\documentclass{iopjournal}

\usepackage{amsmath,amssymb}

\usepackage{braket}
\usepackage{bm}

\usepackage{subcaption}
\usepackage{overpic}
\usepackage{float}
\usepackage{array}
\usepackage{booktabs}

\usepackage{ragged2e}
\usepackage[english]{babel}
\usepackage{csquotes}

\usepackage{soul}
\usepackage[normalem]{ulem}

\usepackage{tikz}

\usepackage[ruled,vlined,linesnumbered]{algorithm2e}

\usepackage{caption}
\newcommand{\lightgreenbox}[1]{%
\begin{tikzpicture}
\node[
  rounded corners=4pt,
  draw=green!45!black,
  line width=1.2pt,
  fill=green!10,
  inner sep=8pt,
  text width=\linewidth-16pt,
  align=left
] {#1};
\end{tikzpicture}%
}

\newcommand{\lightbluebox}[1]{%
\begin{tikzpicture}
\node[
  rounded corners=4pt,
  draw=blue!45!black,
  line width=1.2pt,
  fill=blue!8,
  inner sep=8pt,
  text width=\linewidth-16pt,
  align=left
] {#1};
\end{tikzpicture}%
}

\begin{document}

\articletype{Paper}

%\title{Episodic-Memory-augmented test-time optimization with large language models:\\
%a black-box design benchmark on quantum circuit synthesis}
\title{Quantum Circuit Generation via test-time learning with large language models}

\author{Adriano Macarone-Palmieri$^1$\orcid{0000-0001-6736-3314} and Rosario Lo Franco$^1$\orcid{0000-0002-3281-0935 }}

\affil{$^1$Dipartimento di Ingegneria, Universit\`{a} degli Studi di Palermo, Viale delle Scienze, 90128 Palermo, Italy}

\email{adriano.macaronepalmieri@unipa.it}


\begin{abstract}
Deploying large language models (LLMs) as optimizers for structured scientific design problems requires effective closed-loop refinement under black-box evaluation, where each candidate assessment is computationally expensive and training data is unavailable. We introduce a \emph{memory-augmented test-time optimization} framework that addresses this challenge through three complementary mechanisms: (i)~an episodic memory trace that retains previously high-scoring candidates as explicit context, guiding future proposals; (ii)~score-difference feedback that signals whether each proposal improved or degraded solution quality, providing a dense reward signal at zero additional oracle cost; and (iii)~restart-from-best sampling that escapes local plateaus by re-initializing the search trajectory from the best previously observed candidate. We benchmark this framework on quantum circuit synthesis, a hard combinatorial design problem in which the goal is to maximize the Meyer-Wallach (MW) global entanglement measure---a black-box oracle whose evaluation cost scales exponentially with qubit count, making sample efficiency critical. On 20-qubit circuits (35 gates), our approach reaches $Q(\psi) = 0.99$ without feedback, demonstrating feasibility under the simplest prompting strategy. Scaling to 25 qubits, where naive prompting plateaus near $Q \approx 0.48$, incorporating score-difference feedback and restart-from-best consistently steers the search toward high-entanglement solutions, with multiple runs reaching $Q = 1.0$ within 45 oracle calls. Against a budget-matched random hill-climbing baseline, which stalls at $Q \lesssim 0.29$, our LLM-guided loop achieves substantially superior sample efficiency. An ablation study confirms the respective roles of the components: removing score-difference feedback substantially increases run-to-run variance and produces several low-scoring outcomes, demonstrating that feedback is the primary convergence driver; removing restart-from-best has a comparatively minor effect; and ablating episodic memory collapses the loop to independent sampling---the large-language-monkeys regime---whose exponential oracle cost makes it infeasible at this scale. Structural analysis of the synthesized states reveals that high-MW solutions frequently resemble stabilizer or graph-state constructions, reflecting both properties of the MW metric and the LLM's prior knowledge of quantum structure. Our results establish quantum circuit synthesis as a challenging and well-defined benchmark for LLM-based black-box optimization, and identify episodic memory and evaluator feedback as the indispensable algorithmic levers for sample-efficient inference-time optimization.
\end{abstract}

\section{Introduction}
\label{sec:intro}

Large language models (LLMs) have emerged as powerful generators of structured artifacts---executable code, formal proofs, molecular strings, hardware descriptions---where correctness is an objective criterion, not a matter of style \cite{llm1,llm2,llm5,llm6}. Yet deploying an LLM as a reliable \emph{optimizer} for such artifacts requires more than one-shot generation. The central challenge is closed-loop refinement under \emph{black-box evaluation}: an external oracle scores each candidate, and the model must use those scores to guide its proposals toward better solutions, without gradient access, without a training set, and often within a tight evaluation budget \cite{llm7,llm8,FlamShepherd2022}.

A natural first response is to scale inference-time compute through repeated sampling \cite{adaptive1,adaptive2,adaptive3,adaptive4}. If enough independent candidates are drawn, rare correct solutions appear with non-negligible probability---the so-called large-language-monkeys effect \cite{brown2024monkey,brown2024largelanguagemonkeysscaling}. However, independent sampling discards all information accumulated from previous evaluations. When each oracle call is expensive, throwing away evaluated candidates is a costly waste.

A complementary direction is \emph{test-time learning}: feeding evaluated outcomes back into the prompt so that successive queries benefit from accumulated experience. The Dynamic Cheatsheet framework \cite{suzgun2025dynamiccheatsheettesttimelearning} showed that maintaining a curated textual memory of past outcomes substantially improves solution quality in reasoning tasks. TextGrad \cite{TextGrad} demonstrated that differentiating through natural-language feedback can guide generation in molecule design. These results suggest a general principle: when evaluation is expensive, reusing every assessed candidate as a learning signal is essential.

In this work, we study these ideas on \emph{quantum circuit synthesis}---a combinatorial design problem that makes the black-box optimization challenge concrete and measurable. The task is to find a sequence of quantum gates, drawn from a fixed discrete set, such that the resulting output state maximises the Meyer-Wallach (MW) global entanglement measure \cite{MeyerWallach}. This benchmark has two properties that make it a demanding testbed for LLM-based optimizers: (i)~the evaluation oracle (exact quantum simulation) is costly, scaling exponentially with qubit count, making sample efficiency critical; and (ii)~the LLM carries genuine prior knowledge about quantum computing, creating a realistic setting in which structured generation can be leveraged. Prior ML methods have addressed this problem with diffusion networks producing solutions for up to 9~qubits \cite{llm3}, and reinforcement learning for 4--6~qubits \cite{RLentanglement,RLdisentanglement,Lockwood2020}, but all require either training data or a differentiable simulator. Our approach requires neither.

We introduce a \emph{memory-augmented test-time optimization} framework built from three components, each targeting a specific failure mode of naive LLM querying:
\begin{itemize}
    \item \textbf{Episodic memory.} The best previously evaluated circuit is retained as explicit context in each new prompt, providing the model with a high-quality warm-start rather than querying from a fixed random seed. Without this component the loop collapses to the large-language-monkeys regime \cite{brown2024largelanguagemonkeysscaling}: independent sampling that is economically expensive and sample-inefficient when each oracle call costs $\mathcal{O}(2^n)$ operations.
    \item \textbf{Score-difference feedback.} A textual reward or penalty signal is appended to each prompt, encoding the change in MW score between consecutive proposals ($\Delta Q$). This provides a dense directional cue at zero additional oracle cost, analogous to a textual reward signal in reinforcement learning \cite{Sutton1998}.
    \item \textbf{Restart-from-best sampling.} When a query trajectory stalls, the search is re-initialized from the globally best observed candidate rather than from a new random circuit, replacing expensive restarts with a principled warm start.
\end{itemize}

We evaluate these components on 20- and 25-qubit circuit synthesis using frontier LLMs (GPT\,5.1 and GPT\,5.2). On 20-qubit problems, episodic memory alone achieves $Q(\psi) = 0.99$ in the best case. Scaling to 25 qubits, where naive prompting plateaus near $Q \approx 0.48$, adding feedback and restart-from-best consistently reaches $Q = 1.0$ within 45 oracle calls---far exceeding a budget-matched random hill-climbing baseline that stalls at $Q \lesssim 0.29$. Structural analysis of high-scoring solutions reveals a preference for stabilizer and graph-state topologies, providing interpretable insight into what the LLM has effectively learned to produce.

The paper is organised as follows. Section~\ref{sec:method} presents the framework, the evaluation oracle, the benchmark, and the full algorithm. Section~\ref{sec:results} reports experimental results. Section~\ref{sec:conclusion} discusses implications and future directions.

\section{Methods}
\label{sec:method}

\subsection{Problem formulation}

We cast quantum circuit synthesis as an instance of \emph{black-box LLM optimization}. Formally, let $\mathcal{C}$ denote the space of fixed-length gate sequences over a discrete gate alphabet $\mathcal{G}$, and let $f: \mathcal{C} \to [0,1]$ be a black-box evaluation oracle (Section~\ref{sec:oracle}). At each iteration $i$, an LLM $\mathcal{M}$ receives a prompt $p_i$ constructed from a candidate circuit $\mathcal{U}_i \in \mathcal{C}$ and auxiliary context, and returns a new candidate:
\begin{equation}
    \mathcal{U}_{i+1} = \mathcal{M}(p_i).
\end{equation}
The goal is to maximise $f(\mathcal{U})$ over a fixed oracle-call budget $T \cdot R$, where $T$ is the number of steps per query and $R$ is the number of restarts. This setting differs from standard supervised learning in that (i)~no gradient flows through $f$, (ii)~each call to $f$ may be expensive, and (iii)~no training examples are available beyond the candidates evaluated online during the search. The framework is agnostic to the specific oracle $f$ and extends directly to other design tasks where a black-box evaluator is available.

\subsection{Memory-augmented prompting}
\label{sec:memory}

Our approach is inspired by the \emph{Dynamic Cheatsheet} (DC) strategy \cite{suzgun2025dynamiccheatsheettesttimelearning}, a form of test-time learning in which the model is conditioned on a curated history of previous outcomes. We implement a minimal but effective instantiation of this idea with two components.

\paragraph{Episodic memory.} At each step $i$, the prompt $p_i$ includes the current best circuit $\mathcal{U}^* = \arg\max_{j \leq i} f(\mathcal{U}_j)$ as explicit context. This provides a high-quality warm-start and prevents the model from regressing to low-quality random structures. Crucially, without this component the iterative loop reduces to \emph{independent sampling}: each query starts from a fresh random circuit, accumulates no information about what has worked, and the resulting behaviour is precisely the large-language-monkeys regime \cite{brown2024largelanguagemonkeysscaling}---a strategy that can find solutions in principle, but only by issuing exponentially many oracle calls until a rare high-quality candidate is encountered by chance. When each oracle evaluation scales as $2^n$ in state-vector memory, this cost is prohibitive for $n \geq 20$. Episodic memory is therefore not an optional enhancement but the mechanism that transforms the loop from expensive random search into a directed, sample-efficient optimization.

\paragraph{Score-difference feedback.} The prompt also includes a scalar feedback signal
\begin{equation}
    \Delta Q_i = f(\mathcal{U}_i) - f(\mathcal{U}_{i-1}),
\end{equation}
expressed as a natural-language sentence---for example, \textit{``You obtained an improvement of $+0.12$ in the MW measure''} or \textit{``You made it worse by $-0.05$''}. This signal requires no additional oracle calls and provides the model with directional information about its last proposal.

An output curation step is applied at every iteration: the model output is post-processed to extract a syntactically valid gate-list string, discarding any surrounding explanation, code fences, or non-ASCII characters that would prevent circuit construction.

\subsection{Evaluation oracle: Meyer-Wallach global entanglement}
\label{sec:oracle}

We use the Meyer-Wallach (MW) global entanglement measure \cite{MeyerWallach} as the objective $f$. For a pure $N$-qubit state $\ket{\psi}$, the measure is defined as
\begin{equation}
    Q(\psi) = \frac{4}{N} \sum_{i=1}^N \left( 1 - \mathrm{Tr}\!\left[ \rho_i^2 \right] \right),
    \label{eq:meyer-wallach}
\end{equation}
where $\rho_i = \mathrm{Tr}_{\{1,\ldots,N\}\setminus i}(\ket{\psi}\!\bra{\psi})$ is the reduced density operator of qubit $i$. The MW measure satisfies $0 \leq Q(\psi) \leq 1$, with $Q=1$ for maximally entangled states (e.g., GHZ or products of Bell pairs with non-trivial rotations) and $Q=0$ for fully separable states. It is permutation-invariant and experimentally accessible \cite{MW_operationalInterpretation,Wallach2008}, and its scalar value makes it a natural feedback signal for the prompt. A full derivation and discussion of its properties are provided in Appendix~\ref{app:mw}.

\subsection{Benchmark: quantum circuit synthesis}
\label{sec:benchmark}

We represent circuits as ordered lists of gate tuples acting on $n$ qubits initialised in $\ket{0}^{\otimes n}$. The gate alphabet is restricted to $\mathcal{G} = \{\mathrm{H},\,\mathrm{CNOT},\,\mathrm{RY}\}$, motivated by hardware-native compilations in which multi-qubit gates such as SWAP decompose into three CNOT gates \cite{QuantumCompilation2022,QuantumCompilation2024,QuantumCompilationMoro2021}. To keep the search space tractable, the continuous rotation angle of $\mathrm{RY}$ is discretised to the set $\{3, 10, 25\}$ degrees (standard experiments) or $\{0.1, 0.42, 1.0\}$ radians (non-Clifford experiments; Section~\ref{sec:25q}). The circuit length (number of gates) is fixed throughout each experiment, and the model is explicitly instructed not to add or remove gates---only to replace existing ones. This constraint prevents the degenerate strategy of maximising $Q$ by appending an arbitrary number of CNOT gates, and forces the model to discover efficient entangling topologies under a resource budget.

\begin{figure}
    \centering
\includegraphics[width=0.8\linewidth]{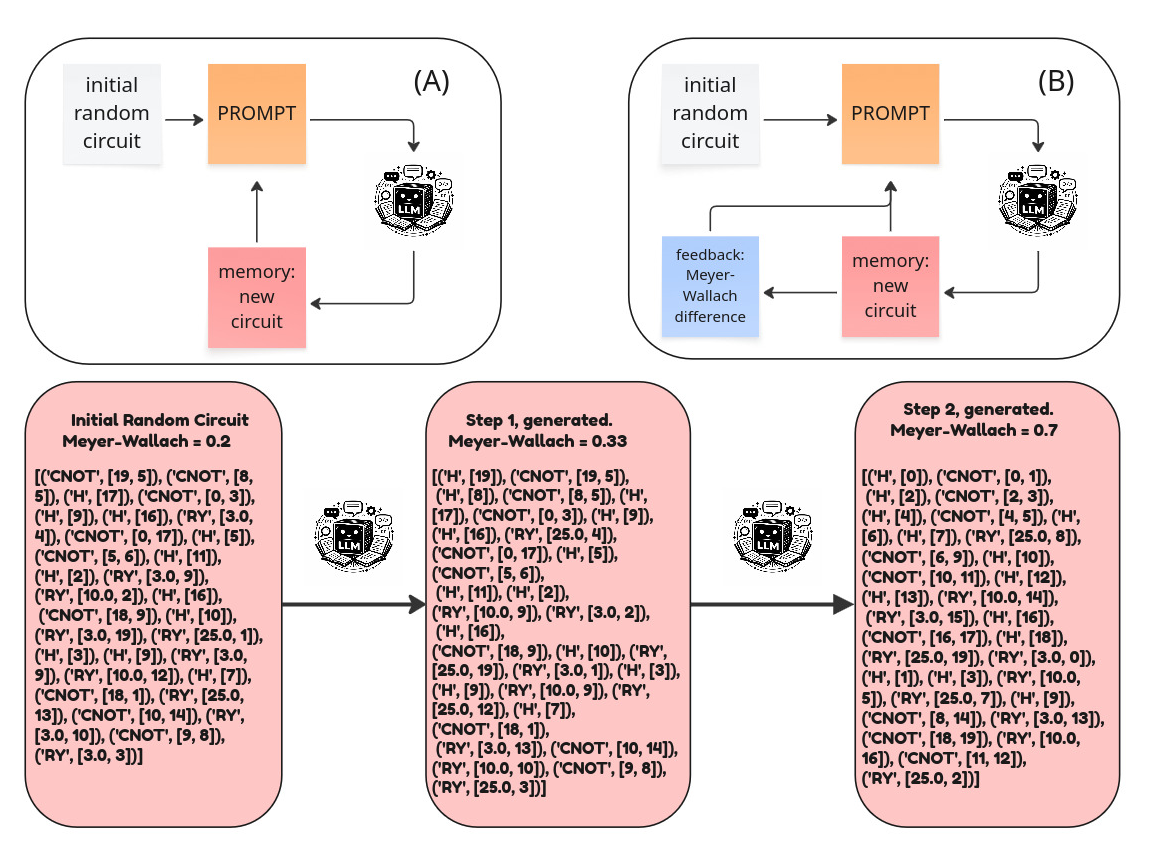}
    \caption{\justifying \textbf{Box (A).} Inference loop without score-difference feedback: the current best circuit is placed in the prompt and the LLM proposes a replacement.
    \textbf{Box (B).} Inference loop with feedback: the prompt additionally includes the MW score change $\Delta Q$ from the previous step, expressed as a natural-language reward or penalty.
    \textbf{Lower-row red boxes.} Example of iterative circuit generation over 6 steps on a small instance, using GPT\,5.1 without feedback. The initial circuit (left) is sampled randomly from $\mathcal{G}$ with angles in $\{3,10,25\}$. The model replaces gates creatively at each step, and the final circuit (right) is a product of Bell pairs with non-trivial $\mathrm{RY}$ rotations.
    }
    \label{fig:train for 6 steps}
\end{figure}

The LLM receives the circuit as a plain-text list of tuples and outputs a modified list in the same format. An example of this representation and of the iterative generation process is shown in Fig.~\ref{fig:train for 6 steps}.

\subsection{Algorithm}
\label{sec:algorithm}

Algorithm~\ref{alg:main} provides a formal description of the full memory-augmented optimization loop.

\begin{algorithm}[t]
\caption{Memory-augmented LLM optimization}
\label{alg:main}
\KwIn{Initial circuit $\mathcal{U}_0 \in \mathcal{C}$, LLM $\mathcal{M}$, oracle $f$, steps per query $T$, restart budget $R$}
\KwOut{Best circuit $\mathcal{U}^*$}
$\mathcal{U}^* \leftarrow \mathcal{U}_0$; $\;Q^* \leftarrow f(\mathcal{U}_0)$\;
\For{$r = 1$ \KwTo $R$}{
    $\mathcal{U}_{\rm curr} \leftarrow \mathcal{U}^*$ \tcp*{restart from global best}
    $Q_{\rm prev} \leftarrow f(\mathcal{U}_{\rm curr})$\;
    \For{$t = 1$ \KwTo $T$}{
        $\Delta Q \leftarrow Q_{\rm prev} - f(\mathcal{U}^*)$ \tcp*{score-difference feedback}
        Construct prompt $p$ from $(\mathcal{U}_{\rm curr},\, \mathcal{U}^*,\, \Delta Q)$\;
        $\mathcal{U}_{\rm new} \leftarrow \mathrm{curate}(\mathcal{M}(p))$ \tcp*{generate and parse}
        $Q_{\rm new} \leftarrow f(\mathcal{U}_{\rm new})$\;
        \If{$Q_{\rm new} > Q^*$}{
            $\mathcal{U}^* \leftarrow \mathcal{U}_{\rm new}$; $\;Q^* \leftarrow Q_{\rm new}$\;
        }
        $\mathcal{U}_{\rm curr} \leftarrow \mathcal{U}_{\rm new}$; $\;Q_{\rm prev} \leftarrow Q_{\rm new}$\;
    }
}
\Return $\mathcal{U}^*$
\end{algorithm}

\subsection{Models and prompting strategy}
\label{sec:models}

We evaluate two frontier models: GPT\,5.1 (used in 20-qubit experiments without feedback) and GPT\,5.2 (used in 25-qubit experiments with the full framework). Both are accessed via API with temperature set to encourage creative exploration, as described in Appendix~\ref{app:prompting}.

For 20-qubit experiments, the prompt places the current circuit in context and asks the model to improve MW entanglement by modifying gates---a minimal test of whether the model can leverage its internal quantum knowledge without any feedback signal (Box~A in Fig.~\ref{fig:train for 6 steps}).

For 25-qubit experiments, we activate the full framework: score-difference feedback (Box~B), episodic memory from the best previous circuit, and restart-from-best across $R=3$ independent queries of $T=15$ steps each, for a total of 45 oracle calls per experiment. A detailed prompt template is provided in Appendix~\ref{app:prompting}.

\section{Experiments}
\label{sec:results}

Across all experiments, we observe that the optimal solutions discovered by the framework gravitate towards stabilizer states \cite{Stabilizers,RevModPhysStabilizers} and graph/cluster states \cite{vandennest2004LCgraph,vandennest2005LUvsLC}---a structural regularity we analyse in Section~\ref{sec:25q} and Appendix~\ref{app:state analysis}. We also observe a dramatic performance gap between frontier and smaller open-weight models; the latter are assessed in Appendix~\ref{app:smallLM} as a model-quality ablation.

\subsection{Experimental setup}
\label{sec:setup}

All circuits are initialised in $\ket{0}^{\otimes n}$ and evaluated by exact state-vector simulation. The gate alphabet is $\mathcal{G} = \{\mathrm{H},\,\mathrm{CNOT},\,\mathrm{RY}\}$. For primary experiments, $\mathrm{RY}$ angles are fixed to $\{3, 10, 25\}$ degrees; Section~\ref{sec:25q} also reports a second set of non-Clifford angles $\{0.1, 0.42, 1.0\}$ radians. Circuit length is held constant throughout each experiment (no gates may be added or removed). GPT\,5.1 is used in 20-qubit experiments; GPT\,5.2 in 25-qubit experiments. The 25-qubit protocol uses $R=3$ restarts of $T=15$ steps each (45 oracle calls total). Success is defined as $Q(\psi) \geq 0.8$ unless otherwise stated.

\subsection{20-qubit circuits: memory without feedback}
\label{sec:20q}

We first evaluate GPT\,5.1 with episodic memory but no score-difference feedback (Box~A in Fig.~\ref{fig:train for 6 steps}), varying circuit length from 25 to 45 gates on $n=20$ qubit registers. This setting probes whether the model's prior knowledge of quantum structure is sufficient to drive improvement without any explicit reward signal.

\begin{table}[ht]
\centering
\caption{MW improvements obtained with GPT\,5.1 using episodic memory without feedback. Each row reports one or more independent runs at a fixed gate count. The model consistently respects the gate-count constraint and improves over the initial circuit in all but one case (40 gates, second run). A single run at 35 gates reaches $Q(\psi)=0.99$, demonstrating that frontier models can discover near-optimal solutions under minimal guidance---but with a success rate of $\approx 10\%$ for $Q(\psi)>0.8$.}
\label{table:20}
\begin{tabular}{ccc}
\toprule
Gates & Initial $Q(\psi)$ & Final $Q(\psi)$ \\
\midrule
25 & 0.23 & 0.70 \\
30 & 0.20, 0.23 & 0.67, 0.70 \\
35 & 0.48 & \textbf{0.99} \\
40 & 0.24, 0.24 & 0.24, 0.43 \\
45 & 0.25, 0.35, 0.22 & 0.38, 0.35, 0.83 \\
\bottomrule
\end{tabular}
\end{table}

Table~\ref{table:20} shows that episodic memory consistently yields improvements, demonstrating that LLMs can discover higher-entanglement circuits beyond what repeated independent sampling would find \cite{brown2024largelanguagemonkeysscaling}. However, the $\approx 10\%$ success rate at $Q(\psi)>0.8$ motivates the additions introduced for the 25-qubit setting.

\subsection{25-qubit 35-gates circuits: feedback and restart-from-best}
\label{sec:25q}

When scaling to $n=25$ qubits, naive prompting (memory only, no feedback) saturates near $Q(\psi) \approx 0.48$ regardless of circuit length. We activate the full framework---score-difference feedback and restart-from-best---and report results across 11 consecutive runs in Table~\ref{tab:25}. This is implemented with the intention to take advantage of the reply variability that advanced models possess \cite{smith2025comprehensiveanalysislargelanguage}, and reuse it as a booster for sampling quality and a workaround against potential "local minima stuck".

\begin{table}[ht]
\centering
\caption{Results on 25-qubit 35-gates circuit synthesis with GPT\,5.2 using score-difference feedback and restart-from-best ($R=3$ queries of $T=15$ steps each). Rows A and C use pre-sampled high-quality starting circuits. Rows B and D1 start from random circuits and stall. Experiments D2 and D3, which restart from the D1 plateau circuit, also fail to improve---suggesting that certain circuit topologies constitute hard plateaus for the model. ``done'' indicates the experiment reached its target before exhausting queries. All 11 runs are reported without selection.}
\label{tab:25}
\begin{tabular}{lccc}
\toprule
Run & Query 1 & Query 2 & Query 3 \\
\midrule
A  & $0.47\!\to\!0.48$ & $0.48\!\to\!0.66$ & $0.66\!\to\!0.8\!\to\!\mathbf{1}$ \\
B  & $0.05\!\to\!0.77$ & no improv. & no improv. \\
C1 & $0.45\!\to\!\mathbf{1}$ & done & done \\
C2 & $0.45\!\to\!0.99$ & done & done \\
D1 & $0.11\!\to\!0.32$ & no improv. & no improv. \\
D2 & $0.32\!\to\!0.32$ & no improv. & no improv. \\
D3 & $0.32\!\to\!0.32$ & no improv. & no improv. \\
E  & $0.22\!\to\!0.92$ & done & done \\
F  & $0.20\!\to\!0.96$ & done & done \\
G  & $0.19\!\to\!0.52$ & no improv. & no improv. \\
H  & $0.16\!\to\!0.98$ & done & done \\
\bottomrule
\end{tabular}
\end{table}

Two patterns emerge from Table~\ref{tab:25}. First, starting circuits with $Q(\psi) \gtrsim 0.45$ consistently lead to high or perfect MW values within a single query (rows A, C, E, F, H)---suggesting a threshold above which the model can reliably close the gap to optimality. Second, certain topologies act as hard plateaus (rows D1--D3): even restarting the search from a plateau circuit cannot escape it, implying a model-level failure that cannot be resolved by additional oracle calls from the same starting point.

The representative solution from experiment~A decomposes as
\begin{equation}
    \Big( \bigotimes_{i=0}^{11}\ket{\Phi^+_{i,\,i+12}} \Big)\otimes \ket{{\rm GHZ}}_{11,23,24},
\end{equation}
a product of 11 Bell pairs and one 3-qubit GHZ state---an entirely Clifford stabilizer state LC-equivalent to a graph state. Additional state analyses are given in Appendix~\ref{app:state analysis}.

The model also discovers non-stabilizer solutions when prompted with $\mathrm{RY}$ angles $\{3, 10, 25\}$ degrees. One example achieving $Q(\psi)=0.99$ is shown below:
\begin{minipage}{\columnwidth}
\lightbluebox{\footnotesize[('H', [0]), ('CNOT', [0, 1]), ('RY', [25.0, 2]), ('CNOT', [1, 2]), ('CNOT', [2, 3]), ('RY', [10.0, 3]), ('H', [4]), ('CNOT', [4, 5]), ('CNOT', [5, 6]), ('CNOT', [6, 7]), ('CNOT', [7, 8]), ('H', [9]), ('CNOT', [9, 10]), ('RY', [25.0, 10]), ('H', [11]), ('CNOT', [10, 11]), ('CNOT', [11, 12]),
('CNOT', [12, 13]), ('CNOT', [13, 14]), ('RY', [10.0, 14]), ('CNOT', [14, 15]), ('CNOT', [15, 16]), ('CNOT', [16, 17]), ('RY', [25.0, 18]), ('CNOT', [17, 18]), ('CNOT', [18, 19]), ('CNOT', [19, 20]), ('RY', [10.0, 20]), ('CNOT', [20, 21]), ('CNOT', [21, 22]), ('CNOT', [22, 23]), ('RY', [25.0, 24]), ('CNOT', [23, 24]), ('CNOT', [8, 16]), ('CNOT', [3, 21]), ('RY', [3.0, 6])].}
\end{minipage}
\vspace{3pt}

\noindent Table~\ref{tab:new angles} reports results with non-Clifford $\mathrm{RY}$ angles $\{0.1, 0.42, 1.0\}$~rad, confirming that the framework generalises beyond the Clifford regime.

\begin{table}[ht]
\centering
\caption{Results with non-Clifford $\mathrm{RY}$ angles $\{0.1, 0.42, 1.0\}$~rad, using GPT\,5.2 with the full framework. Four runs are reported. Notably, in runs B and D the model achieves $Q(\psi)\geq 0.96$ without using $\mathrm{RY}$ gates at all, discovering Clifford solutions purely when non-Clifford gates are available.}
\label{tab:new angles}
\begin{tabular}{lccc}
\toprule
Run & Query 1 & Query 2 & Query 3 \\
\midrule
A & $0.16\!\to\!0.63$ & no improv. & $0.63\!\to\!0.68$ \\
B & $0.20\!\to\!\mathbf{0.96}$ & done & done \\
C & $0.29\!\to\!0.72$ & no improv. & no improv. \\
D & $0.33\!\to\!0.59$ & $0.59\!\to\!0.76$ & $0.76\!\to\!\mathbf{1}$ \\
\bottomrule
\end{tabular}
\label{tab non-Clifford}
\end{table}

\subsection{Ablation study for non-Clifford}

The ablation study serves to properly evaluate the role of each component inside our workflow strategy for circuit synthesis. To better understand how each component works separately, we run an ablation for the 25-qubit 35-gate problem with non-Clifford gate set. Imposing a fixed computational budget of 45 oracle calls, Table~\ref{tab ablation} reports two conditions: the loop without restart-from-best (memory and feedback active), and the loop without score-difference feedback (memory and restart active). This design isolates the contribution of each component while keeping oracle cost constant, and serves to unfold how much a frozen LLM benefits from RL-inspired prompting inside the inference loop versus how much it relies on its internal knowledge alone.

\begin{table}[ht]
\centering
\caption{Ablation study for GPT\,5.2 optimizing the non-Clifford gate set, fixed budget of 45 inference steps. \textbf{No restart}: memory and feedback active, restart-from-best removed. \textbf{No feedback}: memory and restart-from-best active, score-difference feedback removed. Results for the full loop (all three components) are reported in Table~\ref{tab:25} and Table~\ref{tab:new angles}. The \textbf{No memory} condition is not run directly because removing episodic memory causes each query to start from an independent random circuit, collapsing the loop to large-language-monkey sampling \cite{brown2024largelanguagemonkeysscaling}; this regime is already characterised by Table~\ref{table:20} ($\approx 10\%$ success rate for $Q>0.8$) and the random hill-climbing baseline of Section~\ref{sec:baseline} ($Q \lesssim 0.29$), and incurs exponentially higher oracle cost per successful solution. Each row is one independent run (A--J). Removing feedback reduces systematic convergence to high $Q$ values; removing restart-from-best has a smaller impact.}
\label{tab ablation}
\begin{tabular}{ccc}
\toprule
Run & No restart & No feedback \\
\midrule
A & 0.87 & 1.00 \\
B & 1.00 & 0.24 \\
C & 0.83 & 0.84 \\
D & 1.00 & 0.67 \\
E & 1.00 & 1.00 \\
F & 0.96 & 0.92 \\
G & 0.80 & 0.80 \\
H & 0.77 & 0.69 \\
I & 0.94 & 0.43 \\
J & 0.67 & 0.79 \\
\bottomrule
\end{tabular}
\end{table}

The test confirms an expected but not guaranteed trend. Even though we are dealing with frozen model inference, which is mathematically different from a full-fledged reinforcement learning policy optimization, the role of RL-crafted prompting offers a useful piece of information to the model, without making it shift into out-of-distribution regime. The no-feedback condition (Table~\ref{tab ablation}, right column) reveals that removing score-difference feedback increases variance and produces several low-scoring runs (e.g.\ runs B and I drop to $Q=0.24$ and $Q=0.43$ respectively), confirming that feedback is the key algorithmic lever. By contrast, the no-restart condition (left data column) shows that removing restart-from-best incurs a smaller penalty---most runs track closely to the full-loop results of Tables~\ref{tab:25} and~\ref{tab:new angles}---suggesting that restart contributes to robustness but is not strictly necessary when the initial circuit is already of moderate quality. This analysis sheds light on the effect of RL-inspired prompting even when using a frozen model for inference.

A third ablation condition---removing episodic memory entirely---is not reported as a separate column because its behaviour is already characterised elsewhere in the paper. Without memory, each query receives a fresh random circuit as its starting point and the model operates as a pure independent sampler. This is precisely the large-language-monkeys regime \cite{brown2024largelanguagemonkeysscaling}: correct solutions are encountered eventually, but only by accumulating many oracle calls until a rare high-quality draw occurs by chance. The quantitative cost of this regime is visible in two places: (i)~Table~\ref{table:20} shows that even the simplest memory-augmented framework achieves $Q>0.8$ in only $\approx 10\%$ of 20-qubit runs, a rate that would require $\sim 10\times$ more independent draws to match in expectation; (ii)~the random hill-climbing baseline of Section~\ref{sec:baseline}, which is budget-matched but memory-free, plateaus at $Q \lesssim 0.29$ for 25 qubits. The exponential scaling of exact quantum simulation ($\mathcal{O}(2^n)$ memory per oracle call) makes this overhead financially prohibitive: episodic memory is therefore the component that keeps the framework economically viable, converting expensive random search into directed, sample-efficient optimization.

\subsection{Comparison with random hill-climbing baseline}
\label{sec:baseline}

For a fair budget-matched comparison, we implement a classical random-edit hill-climbing baseline using the same gate alphabet $\mathcal{G}$ and the same move set (random gate replacement, wire mutation, and occasional gate swaps). A proposed circuit is accepted only if it improves $Q(\psi)$. The evaluation budget is fixed at 45 oracle calls, identical to the LLM protocol of Table~\ref{tab:25}.

Across ten independent runs starting from the same random initialisation as Table~\ref{tab:25} ($Q(\psi)\approx 0.02$), the baseline fails to exceed $Q(\psi) \approx 0.16$. Starting from a slightly better circuit ($Q(\psi)=0.08$), the best outcome across ten runs reaches $Q(\psi)\approx 0.29$. By contrast, the full LLM framework achieves $Q(\psi) \geq 0.92$ in 7 out of 11 runs under the same budget (Table~\ref{tab:25}). This comparison demonstrates that the LLM's structured generation, conditioned on memory and feedback, provides a substantially more efficient search direction than random local edits within the same oracle budget.

%\noindent{\bf Model quality as a control.} To separate the effect of the optimization loop from the intrinsic capability of the model, we apply the same framework to a smaller reasoning-capable open-weight model for the 25-qubits and 35-gates problems. This model achieves a success rate near $100\%$, but consistently rediscovers only two circuit topologies: nearest-neighbour CNOT ladders and star-topology circuits connecting qubit 0 to all others. This confirms that the loop enables finding solutions regardless of model size, but that the diversity and quality of the discovered solutions scale with the model's reasoning capacity. Full details are in Appendix~\ref{app:smallLM}.

\subsection{Comparison with small model}
\label{sec:small_model}

For the ablation analysis, we focus on the small OSS Hugging Face model only. When the restart is OFF, the model can provide only one single solution, the ladder-type $\mathrm{CNOT}$ connectivity. This is likely a well-known solution that the model samples to solve the problem. When the feedback is provided and the model is prompted to be creative, then the small model can generate just a second solution with connection $\mathrm{CNOT}(0,*)$. This demonstrates two important facts: i) the model's reasoning ability is an important aspect of the problem to find out original solutions, and ii) the protocol by itself can offer an improvement in the sampling ability. As expected, big models can generate more original solutions by leveraging,i.e., bigger reply variety that here translates into novel solutions as the solutions sampled in Appendix \ref{app:state analysis} demonstrate.  

\section{Discussion and Conclusions}
\label{sec:conclusion}

We studied a computationally cheap memory-augmented test-time optimization framework for LLM-guided black-box design and benchmarked it on quantum circuit synthesis under the Meyer-Wallach entanglement metric. The three components---episodic memory, score-difference feedback, and restart-from-best sampling---are individually motivated and jointly necessary: at 25 qubits for a 35 constraint budget of gates, memory alone saturates near $Q \approx 0.48$, while the full framework consistently produces solutions with $Q \geq 0.92$ in the majority of runs, far exceeding a budget-matched random hill-climbing baseline capped at $Q \lesssim 0.29$.

Of the three components, episodic memory plays a qualitatively distinct role. Feedback and restart-from-best modulate \emph{how efficiently} the search converges; episodic memory determines \emph{whether} a directed search exists at all. Without it, every query starts from an independent random circuit and the loop degenerates into the large-language-monkeys regime \cite{brown2024largelanguagemonkeysscaling}---repeated unguided sampling that can in principle find solutions, but only by paying an exponentially larger oracle budget. Since each oracle evaluation costs $\mathcal{O}(2^n)$ in memory and time, this budget is financially and computationally prohibitive for $n \geq 20$. Episodic memory is therefore the mechanism that makes the framework economically viable: it transforms independent, expensive sampling into a cheap directed search by conditioning every new proposal on the best outcome seen so far.

%%%%%Generalizability.
Although our experiments target one specific oracle (MW entanglement) and one domain (quantum circuits), the framework is domain-agnostic. The three components---memory, feedback, restart---are applicable wherever (i)~a black-box scalar evaluator is available, (ii)~structured generation in a fixed format is required, and (iii)~oracle calls are expensive enough that sample efficiency matters. 

%%%%%%%%The role of model quality.
The model-quality ablation (Appendix~\ref{app:smallLM}) reveals that smaller open-weight models can find valid solutions but converge to a narrow set of topologies, while frontier models (GPT\,5.1/5.2) discover qualitatively diverse high-entanglement structures. This suggests that reasoning capacity, not just the loop architecture, is a bottleneck for solution diversity to tackle the broad and more challenging problem of out-of-distribution generalization \cite{miller20-ood1,Hendrycks2020TheMF-ood2,ribeiro-etal-2020-beyond-ood3,pmlr-v139-koh21a-ood4}. As capable open-weight models become more widely available, the framework should extend naturally to fully reproducible, locally hosted pipelines.

%%%%%%%%Limitations.
Interestingly, some circuit topologies act as hard plateaus: even restarting from such circuits fails to escape them, and the mechanism behind this failure is not yet understood. It may reflect a systematic gap in the model's quantum-circuit prior---specific entangling patterns that the model does not associate with high MW values. Fine-tuning on domain-specific circuit data, or incorporating richer theoretical knowledge as in-context examples, could address this. 

%%%%%%%%%%%Future directions.
From the quantum perspective, future work will explore connectivity-constrained circuit synthesis, hardware-noise-aware objectives, and multi-objective metrics beyond global entanglement. From the ML perspective, the incorporation of cumulative multi-step memory \cite{suzgun2025dynamiccheatsheettesttimelearning}, or RL-based approaches \cite{ttt-discover2026} ---where certain candidate solutions are systematically unimprovable by the model--- is potentially a general feature of LLM-guided optimization that warrants dedicated study.

\ack{A.M.P. and R.L.F. acknowledge support by MUR (Ministero dell'Universit\`{a} e della Ricerca) through the PNRR Project ICON-Q --- Partenariato Esteso NQSTI --- PE00000023 --- Spoke 2 --- CUP: J13C22000680006.}

\funding{This work was supported by the Italian Ministry of University and Research (MUR) through the PNRR Project ICON-Q --- Partenariato Esteso NQSTI --- PE00000023 --- Spoke 2 --- CUP: J13C22000680006.}

\roles{A.M.P.: Conceptualization, Methodology, Software, Investigation, Formal Analysis, Writing -- Original Draft. R.L.F.: Conceptualization, Supervision, Writing -- Review \& Editing.}

\data{The code and prompt templates used in this study are available from the corresponding author upon reasonable request.}

\section*{References}
\bibliographystyle{unsrt}
\bibliography{biblio}

@misc{llm5,
      title={A Survey of Large Language Models}, 
      author={Wayne Xin Zhao and Kun Zhou and Junyi Li and Tianyi Tang and Xiaolei Wang and Yupeng Hou and Yingqian Min and Beichen Zhang and Junjie Zhang and Zican Dong and Yifan Du and Chen Yang and Yushuo Chen and Zhipeng Chen and Jinhao Jiang and Ruiyang Ren and Yifan Li and Xinyu Tang and Zikang Liu and Peiyu Liu and Jian-Yun Nie and Ji-Rong Wen},
      year={2025},
      eprint={2303.18223},
      archivePrefix={arXiv},
      primaryClass={cs.CL},
      url={https://arxiv.org/abs/2303.18223}, 
}

@INPROCEEDINGS{llm6,
  author={Jern, Linus and Uotila, Valter and Yu, Cong and Zhao, Bo},
  booktitle={2025 IEEE International Conference on Quantum Computing and Engineering (QCE)}, 
  title={Agent-Q: Fine-Tuning Large Language Models for Quantum Circuit Generation and Optimization}, 
  year={2025},
  volume={01},
  number={},
  pages={1621-1632},
  doi={10.1109/QCE65121.2025.00179}}

@misc{llm7,
      title={The generative quantum eigensolver (GQE) and its application for ground state search}, 
      author={Kouhei Nakaji and Lasse Bjørn Kristensen and Ryota Kemmoku and Jorge A. Campos-Gonzalez-Angulo and Mohammad Ghazi Vakili and Haozhe Huang and Mohsen Bagherimehrab and Christoph Gorgulla and FuTe Wong and Alex McCaskey and Jin-Sung Kim and Thien Nguyen and Pooja Rao and Qi Gao and Michihiko Sugawara and Naoki Yamamoto and Alán Aspuru-Guzik},
      year={2025},
      eprint={2401.09253},
      archivePrefix={arXiv},
      primaryClass={quant-ph},
      url={https://arxiv.org/abs/2401.09253}, 
}

@article{llm1,
author = {Chang, Yupeng and Wang, Xu and Wang, Jindong and Wu, Yuan and Yang, Linyi and Zhu, Kaijie and Chen, Hao and Yi, Xiaoyuan and Wang, Cunxiang and Wang, Yidong and Ye, Wei and Zhang, Yue and Chang, Yi and Yu, Philip S. and Yang, Qiang and Xie, Xing},
title = {A Survey on Evaluation of Large Language Models},
year = {2024},
issue_date = {June 2024},
publisher = {Association for Computing Machinery},
address = {New York, NY, USA},
volume = {15},
number = {3},
issn = {2157-6904},
url = {https://doi.org/10.1145/3641289},
doi = {10.1145/3641289},
journal = {ACM Trans. Intell. Syst. Technol.},
month = mar,
articleno = {39},
numpages = {45}
}

@article{llm2,
author = {Shanahan, Murray},
title = {Talking about Large Language Models},
year = {2024},
issue_date = {February 2024},
publisher = {Association for Computing Machinery},
address = {New York, NY, USA},
volume = {67},
number = {2},
issn = {0001-0782},
url = {https://doi.org/10.1145/3624724},
doi = {10.1145/3624724},
journal = {Commun. ACM},
month = jan,
pages = {68–79},
numpages = {12}
}

@article{llm3,
  title = {Quantum circuit synthesis with diffusion models},
  volume = {6},
  ISSN = {2522-5839},
  url = {http://dx.doi.org/10.1038/s42256-024-00831-9},
  DOI = {10.1038/s42256-024-00831-9},
  pages = {5},
  journal = {Nat. Mach. Intell.},
  publisher = {Springer Science and Business Media LLC},
  author = {F\"{u}rrutter,  Florian and Muñoz-Gil,  Gorka and Briegel,  Hans J.},
  year = {2024},
  month = may,
  pages = {515–524}
}

@article{FlamShepherd2022,
  title = {Language models can learn complex molecular distributions},
  volume = {13},
  pages = {3293},
  journal = {Nat. Comm.},
  publisher = {Springer Science and Business Media LLC},
  author = {Flam-Shepherd,  Daniel and Zhu,  Kevin and Aspuru-Guzik,  Alán},
    year = {2022},
    url = {http://dx.doi.org/10.1038/s41467-022-30839-x},
  DOI = {10.1038/s41467-022-30839-x},
}

@book{Sutton1998,
  author = {Sutton, Richard S. and Barto, Andrew G.},
  edition = {Second},
  publisher = {The MIT Press},
  title = {Reinforcement Learning: An Introduction},
  url = {http://incompleteideas.net/book/the-book-2nd.html},
  year = {2018 }
}

@misc{brown2024largelanguagemonkeysscaling,
      title={Large Language Monkeys: Scaling Inference Compute with Repeated Sampling}, 
      author={Bradley Brown and Jordan Juravsky and Ryan Ehrlich and Ronald Clark and Quoc V. Le and Christopher Ré and Azalia Mirhoseini},
      year={2024},
      eprint={2407.21787},
      archivePrefix={arXiv},
      primaryClass={cs.LG},
      url={https://arxiv.org/abs/2407.21787}, 
}

@misc{suzgun2025dynamiccheatsheettesttimelearning,
      title={Dynamic Cheatsheet: Test-Time Learning with Adaptive Memory}, 
      author={Mirac Suzgun and Mert Yuksekgonul and Federico Bianchi and Dan Jurafsky and James Zou},
      year={2025},
      eprint={2504.07952},
      archivePrefix={arXiv},
      primaryClass={cs.LG},
      url={https://arxiv.org/abs/2504.07952}, 
}

@article{RLentanglement,
  title = {Reinforcement-learning generation of four-qubit entangled states},
  author = {Giordano, Sara and Martin-Delgado, Miguel A.},
  journal = {Phys. Rev. Res.},
  volume = {4},
  issue = {4},
  pages = {043056},
  numpages = {18},
  year = {2022},
  month = {Oct},
  publisher = {American Physical Society},
  doi = {10.1103/PhysRevResearch.4.043056},
  url = {https://link.aps.org/doi/10.1103/PhysRevResearch.4.043056}
}

@misc{RLdisentanglement,
      title={Reinforcement Learning to Disentangle Multiqubit Quantum States from Partial Observations}, 
      author={Pavel Tashev and Stefan Petrov and Friederike Metz and Marin Bukov},
      year={2024},
      eprint={2406.07884},
      archivePrefix={arXiv},
      primaryClass={quant-ph},
      url={https://arxiv.org/abs/2406.07884}, 
}

@article{Lockwood2020,
  title = {Reinforcement Learning with Quantum Variational Circuit},
  volume = {16},
  ISSN = {2326-909X},
  url = {http://dx.doi.org/10.1609/aiide.v16i1.7437},
  DOI = {10.1609/aiide.v16i1.7437},
  number = {1},
  journal = {Proceedings of the AAAI Conference on Artificial Intelligence and Interactive Digital Entertainment},
  publisher = {Association for the Advancement of Artificial Intelligence (AAAI)},
  author = {Lockwood,  Owen and Si,  Mei},
  year = {2020},
  month = oct,
  pages = {245–251}
}

@article{TextGrad,
  title = {Optimizing generative AI by backpropagating language model feedback},
  volume = {639},
  ISSN = {1476-4687},
  url = {http://dx.doi.org/10.1038/s41586-025-08661-4},
  DOI = {10.1038/s41586-025-08661-4},
  number = {8055},
  journal = {Nature},
  publisher = {Springer Science and Business Media LLC},
  author = {Yuksekgonul,  Mert and Bianchi,  Federico and Boen,  Joseph and Liu,  Sheng and Lu,  Pan and Huang,  Zhi and Guestrin,  Carlos and Zou,  James},
  year = {2025},
  month = mar,
  pages = {609–616}
}

@article{adaptive1,
  title        = {Adaptive Inference-Time Compute: LLMs Can Predict if They Can Do Better, Even Mid-Generation},
  author       = {Rohin Manvi and others},
  year         = {2024},
  journal      = {arXiv preprint arXiv:2410.02725},
  url          = {https://arxiv.org/abs/2410.02725}
}

@techreport{adaptive2,
  title        = {Inference-Time Scaling for Complex Tasks: Where We Stand and What Lies Ahead},
  author       = {{Microsoft Research}},
  institution  = {Microsoft Research},
  year         = {2025},
  url          = {https://www.microsoft.com/en-us/research/wp-content/uploads/2025/03/Inference-Time-Scaling-for-Complex-Tasks-Where-We-Stand-and-What-Lies-Ahead.pdf}
}

@misc{adaptive3,
      title={Inference-Time Computations for LLM Reasoning and Planning: A Benchmark and Insights}, 
      author={Shubham Parashar and Blake Olson and Sambhav Khurana and Eric Li and Hongyi Ling and James Caverlee and Shuiwang Ji},
      year={2025},
      eprint={2502.12521},
      archivePrefix={arXiv},
      primaryClass={cs.AI},
      url={https://arxiv.org/abs/2502.12521}, 
}

@misc{adaptive4,
  title        = {Inference-Time Computation Techniques: Overview and Applications},
  author       = {{Emergent Mind AI Research}},
  year         = {2025},
  howpublished = {\url{https://www.emergentmind.com/topics/inference-time-computation}},
  note         = {Accessed January 2026}
}

@misc{llm8,
      title={Optimizing Ansatz Design in Quantum Generative Adversarial Networks Using Large Language Models}, 
      author={Kento Ueda and Atsushi Matsuo},
      year={2025},
      eprint={2503.12884},
      archivePrefix={arXiv},
      primaryClass={quant-ph},
      url={https://arxiv.org/abs/2503.12884}, 
}

@article{MeyerWallach,
    author = {Meyer, David A. and Wallach, Nolan R.},
    title = {Global entanglement in multiparticle systems},
    journal = {J. Math. Phys.},
    volume = {43},
    number = {9},
    pages = {4273-4278},
    year = {2002},
    month = {09},
    issn = {0022-2488},
    doi = {10.1063/1.1497700},
    url = {https://doi.org/10.1063/1.1497700},
}

@Inbook{Wallach2008,
author="Wallach, Nolan R.",
editor="Tarabusi, Enrico Casadio
and D'Agnolo, Andrea
and Picardello, Massimo",
title="Quantum Computing and Entanglement for Mathematicians",
bookTitle="Representation Theory and Complex Analysis: Lectures given at the C.I.M.E. Summer School held in Venice, Italy June 10--17, 2004",
year="2008",
publisher="Springer Berlin Heidelberg",
address="Berlin, Heidelberg",
pages="345--376",
isbn="978-3-540-76892-0",
doi="10.1007/978-3-540-76892-0_6",
url="https://doi.org/10.1007/978-3-540-76892-0_6"
}

@article{MW_operationalInterpretation,
  title = {Operational Interpretation for Global Multipartite Entanglement},
  author = {Boixo, S. and Monras, A.},
  journal = {Phys. Rev. Lett.},
  volume = {100},
  issue = {10},
  pages = {100503},
  numpages = {4},
  year = {2008},
  month = {Mar},
  publisher = {American Physical Society},
  doi = {10.1103/PhysRevLett.100.100503},
  url = {https://link.aps.org/doi/10.1103/PhysRevLett.100.100503}
}

@inbook{QuantumCompilation2022,
  title = {Quantum Compiling},
  ISBN = {9783030897468},
  url = {http://dx.doi.org/10.1007/978-3-030-89746-8_2},
  DOI = {10.1007/978-3-030-89746-8_2},

  booktitle = {Quantum Computing Environments},
  publisher = {Springer International Publishing},
  author = {Maronese,  Marco and Moro,  Lorenzo and Rocutto,  Lorenzo and Prati,  Enrico},
  year = {2022},
  pages = {39–74}
}

@InProceedings{QuantumCompilation2024,
author="Cardama, F. Javier
and V{\'a}zquez-P{\'e}rez, Jorge
and Pena, Tom{\'a}s F.
and Pichel, Juan C.
and G{\'o}mez, Andr{\'e}s",
editor="Caino-Lores, Silvina
and Zeinalipour, Demetris
and Doudali, Thaleia Dimitra
and Singh, David E.
and Garz{\'o}n, Gracia Ester Mart{\'i}n
and Sousa, Leonel
and Andrade, Diego
and Cucinotta, Tommaso
and D'Ambrosio, Donato
and Diehl, Patrick
and Dolz, Manuel F.
and Jukan, Admela
and Montella, Raffaele
and Nardelli, Matteo
and Garcia-Gasulla, Marta
and Neuwirth, Sarah",
title="Quantum Compilation Process: A Survey",
booktitle="Euro-Par 2024: Parallel Processing Workshops",
year="2025",
publisher="Springer Nature Switzerland",
address="Cham",
pages="100--112",

}

@article{QuantumCompilationMoro2021,
  title = {Quantum compiling by deep reinforcement learning},
  volume = {4},
  ISSN = {2399-3650},

  url = {http://dx.doi.org/10.1038/s42005-021-00684-3},
  DOI = {10.1038/s42005-021-00684-3},
  pages = {178},
  journal = {Commun. Phys.},
  publisher = {Springer Science and Business Media LLC},
  author = {Moro,  Lorenzo and Paris,  Matteo G. A. and Restelli,  Marcello and Prati,  Enrico},
  year = {2021},
  month = aug 
}

@misc{huggingfacestransformersstateoftheartnatural,
      title={HuggingFace's Transformers: State-of-the-art Natural Language Processing}, 
      author={Thomas Wolf and Lysandre Debut and Victor Sanh and Julien Chaumond and Clement Delangue and Anthony Moi and Pierric Cistac and Tim Rault and Rémi Louf and Morgan Funtowicz and Joe Davison and Sam Shleifer and Patrick von Platen and Clara Ma and Yacine Jernite and Julien Plu and Canwen Xu and Teven Le Scao and Sylvain Gugger and Mariama Drame and Quentin Lhoest and Alexander M. Rush},
      year={2020},
      eprint={1910.03771},
      archivePrefix={arXiv},
      primaryClass={cs.CL},
      url={https://arxiv.org/abs/1910.03771}, 
}

@article{Stabilizers,
  title = {Stabilizer Formalism for Operator Quantum Error Correction},
  author = {Poulin, David},
  journal = {Phys. Rev. Lett.},
  volume = {95},
  issue = {23},
  pages = {230504},
  numpages = {4},
  year = {2005},
  month = {Dec},
  publisher = {American Physical Society},
  doi = {10.1103/PhysRevLett.95.230504},
  url = {https://link.aps.org/doi/10.1103/PhysRevLett.95.230504}
}

@article{RevModPhysStabilizers,
  title = {Colloquium: Protecting quantum information against environmental noise},
  author = {Suter, Dieter and \'Alvarez, Gonzalo A.},
  journal = {Rev. Mod. Phys.},
  volume = {88},
  issue = {4},
  pages = {041001},
  numpages = {23},
  year = {2016},
  month = {Oct},
  publisher = {American Physical Society},
  doi = {10.1103/RevModPhys.88.041001},
  url = {https://link.aps.org/doi/10.1103/RevModPhys.88.041001}
}

@article{vandennest2004LCgraph,
  title        = {Efficient algorithm to recognize the local Clifford equivalence of graph states},
  author       = {M. Van den Nest and J. Dehaene and B. De Moor},
  journal      = {Phys. Rev. A},
  volume       = {70},
  pages        = {034302},
  year         = {2004},
  publisher    = {American Physical Society},
  doi          = {10.1103/PhysRevA.70.034302},
  url          = {https://journals.aps.org/pra/abstract/10.1103/PhysRevA.70.034302}
}

@article{vandennest2005LUvsLC,
  title        = {Local unitary versus local Clifford equivalence of stabilizer states},
  author       = {M. Van den Nest and J. Dehaene and B. De Moor},
  journal      = {Phys. Rev. A},
  volume       = {71},
  pages        = {062323},
  year         = {2005},
  publisher    = {American Physical Society},
  doi          = {10.1103/PhysRevA.71.062323},
  url          = {https://journals.aps.org/pra/abstract/10.1103/PhysRevA.71.062323} 
}

@misc{smith2025comprehensiveanalysislargelanguage,
      title={A Comprehensive Analysis of Large Language Model Outputs: Similarity, Diversity, and Bias}, 
      author={Brandon Smith and Mohamed Reda Bouadjenek and Tahsin Alamgir Kheya and Phillip Dawson and Sunil Aryal},
      year={2025},
      eprint={2505.09056},
      archivePrefix={arXiv},
      primaryClass={cs.CL},
      url={https://arxiv.org/abs/2505.09056}, 
}

@InProceedings{miller20-ood1,
  title = 	 {The Effect of Natural Distribution Shift on Question Answering Models},
  author =       {Miller, John and Krauth, Karl and Recht, Benjamin and Schmidt, Ludwig},
  booktitle = 	 {Proceedings of the 37th International Conference on Machine Learning},
  pages = 	 {6905--6916},
  year = 	 {2020},
  editor = 	 {III, Hal Daumé and Singh, Aarti},
  volume = 	 {119},
  series = 	 {Proceedings of Machine Learning Research},
  month = 	 {13--18 Jul},
  publisher =    {PMLR},
  url = 	 {https://proceedings.mlr.press/v119/miller20a.html}
}

@article{Hendrycks2020TheMF-ood2,
  title={The Many Faces of Robustness: A Critical Analysis of Out-of-Distribution Generalization},
  author={Dan Hendrycks and Steven Basart and Norman Mu and Saurav Kadavath and Frank Wang and Evan Dorundo and Rahul Desai and Tyler Lixuan Zhu and Samyak Parajuli and Mike Guo and Dawn Xiaodong Song and Jacob Steinhardt and Justin Gilmer},
  journal={2021 IEEE/CVF International Conference on Computer Vision (ICCV)},
  year={2020},
  pages={8320-8329},
  url={https://api.semanticscholar.org/CorpusID:220250257}
}

@inproceedings{ribeiro-etal-2020-beyond-ood3,
    title = "Beyond Accuracy: Behavioral Testing of {NLP} Models with {C}heck{L}ist",
    author = "Ribeiro, Marco Tulio  and
      Wu, Tongshuang  and
      Guestrin, Carlos  and
      Singh, Sameer",
    editor = "Jurafsky, Dan  and
      Chai, Joyce  and
      Schluter, Natalie  and
      Tetreault, Joel",
    booktitle = "Proceedings of the 58th Annual Meeting of the Association for Computational Linguistics",
    month = jul,
    year = "2020",
    address = "Online",
    publisher = "Association for Computational Linguistics",
    url = "https://aclanthology.org/2020.acl-main.442/",
    doi = "10.18653/v1/2020.acl-main.442",
    pages = "4902--4912"
}

@InProceedings{pmlr-v139-koh21a-ood4,
  title = 	 {WILDS: A Benchmark of in-the-Wild Distribution Shifts},
  author =       {Koh, Pang Wei and Sagawa, Shiori and Marklund, Henrik and Xie, Sang Michael and Zhang, Marvin and Balsubramani, Akshay and Hu, Weihua and Yasunaga, Michihiro and Phillips, Richard Lanas and Gao, Irena and Lee, Tony and David, Etienne and Stavness, Ian and Guo, Wei and Earnshaw, Berton and Haque, Imran and Beery, Sara M and Leskovec, Jure and Kundaje, Anshul and Pierson, Emma and Levine, Sergey and Finn, Chelsea and Liang, Percy},
  booktitle = 	 {Proceedings of the 38th International Conference on Machine Learning},
  pages = 	 {5637--5664},
  year = 	 {2021},
  editor = 	 {Meila, Marina and Zhang, Tong},
  volume = 	 {139},
  series = 	 {Proceedings of Machine Learning Research},
  month = 	 {18--24 Jul},
  publisher =    {PMLR},
  url = 	 {https://proceedings.mlr.press/v139/koh21a.html}
}

@article{ttt-discover2026,
  title   = {Learning to Discover at Test Time},
  author  = {Yuksekgonul, Mert and Koceja, Daniel and Li, Xinhao 
             and Bianchi, Federico and McCaleb, Jed and Wang, Xiaolong 
             and Kautz, Jan and Choi, Yejin and Zou, James 
             and Guestrin, Carlos and Sun, Yu},
  journal = {arXiv preprint},
  year    = {2026}
}

\appendix

\section{Prompting as an interface for large language models}
\label{app:prompting}

Large language models (LLMs) are conditional generative systems: given an input sequence (the \emph{prompt}), the model produces a continuation sampled from a learned distribution. Prompting therefore acts as a high-leverage interface between user intent and model behaviour that can be iterated rapidly, audited as a textual artefact, and adapted per task without retraining.

\paragraph{Core elements of effective prompts.}
Across diverse tasks, effective prompts include three ingredients: (1)~\emph{goal specification} (what constitutes a correct or useful output), (2)~\emph{context provision} (relevant background, definitions, data, or examples), and (3)~\emph{output contracts} (a declared format and acceptance criteria). Output contracts are particularly important for downstream automation, enabling deterministic parsing and evaluation.

\paragraph{Prompting for robustness.}
Prompt sensitivity motivates systematic evaluation. Robust prompts are those whose performance degrades gracefully under paraphrase, minor perturbations to context, or changes in user phrasing. We recommend testing prompts over representative input distributions and reporting exact prompt templates alongside decoding parameters to support reproducibility.

\lightbluebox{
\texttt{def prompt(gates, feedback, set\_of\_gates):}\\
    \textbf{system} = (
        "You are an expert in PennyLane circuits and entanglement. "\\
        "Modify each tuple using only gates from \{set\_of\_gates\}. "\\
        "Use angles from \{3.0,10.0,25.0\} for the RY gate. "\\
        "Use ASCII only. "\\
        "The evaluation metric is the Meyer-Wallach global entanglement."
    )\\
    \textbf{user} = (
        "You are given a quantum circuit as a list of tuples. "\\
        "GOAL: improve the Meyer-Wallach entanglement of \{gates\} with feedback: \{feedback\}. "\\
        "Transform the circuit substantially---aim for creative leaps, not minor edits. "\\
        "Do NOT add explanations or code fences. "\\
        "Output a valid LIST between <python> and </python>. "\\
        "IMPORTANT: do not add or remove gates, only modify existing ones."
    )}

\paragraph{Prompt design rationale.}
Our system prompt frames the model as an entanglement engineer, constraining its role and reducing off-topic output. The user prompt provides: (i)~the input circuit, (ii)~the metric under optimisation, (iii)~the gate representation format, (iv)~what the model must not do (add/remove gates), and (v)~the output contract (a valid Python list). Adherence to the output contract is essential for the curation step that extracts valid gate lists from model output.

\section{Comparison with smaller open-weight models}
\label{app:smallLM}

To assess the contribution of model scale and reasoning quality independently of the loop architecture, we apply the same episodic memory framework to two smaller open-weight models: \texttt{DeepSeek-R1-Distill-Llama-8B} and \texttt{Zephyr-7B-beta} \cite{huggingfacestransformersstateoftheartnatural}, both accessed via the Hugging Face API at temperature~0.7.

\paragraph{Setup.}
We use a simpler benchmark: $n=4$--5 qubit circuits with 10--20 gates, where random sampling can already achieve $Q(\psi) > 0.8$ with moderate effort (providing an easy oracle as sanity check).

\paragraph{Results.}
The Llama model achieves a mean gain of $0.30 \pm 0.20$ from any starting value; the best observed run improves from $Q=0.70$ to $Q=0.94$. Zephyr achieves $0.20 \pm 0.15$, with best run $0.40 \to 0.50$. Both models improve over the initial circuit, confirming that the loop is effective even with smaller models.

However, both models exhibit systematic failure modes absent in frontier models: (i)~they frequently violate the gate-count constraint, exceeding the maximum gate budget in more than 60\% of queries; (ii)~they rediscover the degenerate strategy of appending many CNOT gates to maximise entanglement---a strategy that is technically correct but that we explicitly forbid in the prompt, indicating insufficient instruction-following fidelity. In contrast, frontier models respect the constraint reliably.

This confirms that the loop architecture is not the bottleneck: the quality, diversity, and constraint-adherence of discovered solutions scales directly with model reasoning capacity.

\section{Characterisation of discovered solutions}
\label{app:state analysis}

We analyse the mathematical structure of representative solutions to understand what the LLM has effectively learned to produce.

\paragraph{MW~$= 0.8$ solution (25 qubits).}
The following circuit, produced by GPT\,5.2 with feedback prompting, achieves $Q(\psi)=0.80$:

\lightgreenbox{[('H', [0]), ('CNOT', [0, 1]), ('H', [2]), ('CNOT', [2, 3]), ('CNOT', [1, 4]), ('H', [5]), ('H', [6]), ('CNOT', [3, 7]), ('CNOT', [4, 8]), ('H', [9]), ('CNOT', [5, 10]), ('CNOT', [6, 11]), ('H', [12]), ('CNOT', [7, 13]), ('CNOT', [8, 14]), ('H', [15]), ('CNOT', [10, 16]), ('CNOT', [11, 17]), ('H', [18]), ('CNOT', [13, 19]), ('CNOT', [14, 20]), ('H', [21]), ('CNOT', [16, 22]), ('CNOT', [17, 23]), ('CNOT', [19, 24])]}

The output state decomposes as
\begin{equation}
\begin{aligned}
\ket{\psi} &=
\frac{\ket{0}^{\otimes 6}+\ket{1}^{\otimes 6}}{\sqrt{2}}_{\{0,1,4,8,14,20\}}\otimes
\frac{\ket{0}^{\otimes 6}+\ket{1}^{\otimes 6}}{\sqrt{2}}_{\{2,3,7,13,19,24\}}\otimes\\
&\frac{\ket{0}^{\otimes 4}+\ket{1}^{\otimes 4}}{\sqrt{2}}_{\{5,10,16,22\}}\otimes
\frac{\ket{0}^{\otimes 4}+\ket{1}^{\otimes 4}}{\sqrt{2}}_{\{6,11,17,23\}}\otimes
\bigotimes_{q\in\{9,12,15,18,21\}} \ket{+}_q .
\end{aligned}
\end{equation}
This is a disconnected stabilizer (graph) state. Its MW is below 1 because the five $\ket{+}$ qubits contribute zero entanglement entropy.

\paragraph{MW~$= 0.66$ solution (25 qubits).}
The following circuit achieves $Q(\psi)=0.66$:

\lightgreenbox{[('H', [0]), ('CNOT', [0, 12]), ('RY', [10.0, 6]), ('CNOT', [6, 18]), ('CNOT', [12, 24]), ('RY', [10.0, 9]), ('H', [1]), ('CNOT', [1, 13]), ('CNOT', [13, 2]), ('CNOT', [2, 14]), ('CNOT', [14, 3]), ('H', [4]), ('H', [5]), ('RY', [10.0, 7]), ('H', [8]), ('CNOT', [8, 20]), ('H', [10]), ('CNOT', [10, 22]), ('H', [11]), ('CNOT', [11, 23]), ('RY', [10.0, 15]), ('H', [16]), ('CNOT', [16, 17]), ('RY', [10.0, 19])].}

This state is a highly disconnected combination of GHZ states, Bell pairs, non-Clifford entangled pairs, and single-qubit $\ket{+}$ and $R_y(\theta)\ket{0}$ states. Qubits 6 and 18 have reduced purity $\approx 0.86$, reflecting a non-stabilizer entangled pair produced by the $\mathrm{RY}(10)$ gate. The state is not a graph state, nor local-Clifford equivalent to one.

\paragraph{General observation.}
Across all high-MW solutions, the model gravitates toward either (i)~products of GHZ states of varying size (maximising each component's contribution to the MW average), or (ii)~highly connected Clifford circuits whose outputs are LC-equivalent to graph states. The preference for Clifford structure is consistent with the LLM's training distribution, which over-represents standard quantum computing textbook circuits.

\section{Meyer-Wallach measure: derivation and properties}
\label{app:mw}

For completeness we provide the full definition and key properties of the Meyer-Wallach measure used as the oracle throughout this work.

Given a pure state $\ket{\psi}$ of an $N$-qubit register in $\mathcal{H} = (\mathbb{C}^2)^{\otimes N}$, define the linear map $\iota_j(z): (\mathbb{C}^2)^{\otimes N} \to (\mathbb{C}^2)^{\otimes(N-1)}$ that restricts qubit $j$ to the computational basis state $z \in \{0,1\}$. The MW measure is then
\begin{equation}
    Q(\psi) = \frac{4}{N} \sum_{j=1}^{N} D\!\left(\iota_j(0)\ket{\psi},\,\iota_j(1)\ket{\psi}\right),
\end{equation}
where $D(u,v) = \|u\|^2\|v\|^2 - |\langle u|v\rangle|^2$ is the generalized concurrence. This definition is equivalent to Eq.~\eqref{eq:meyer-wallach} via $Q(\psi) = \frac{4}{N}\sum_j(1-\mathrm{Tr}[\rho_j^2])$, where $\rho_j$ is the single-qubit reduced density operator \cite{Wallach2008}. Key properties:
\begin{itemize}
    \item $0 \leq Q(\psi) \leq 1$, with $Q=0$ iff $\ket{\psi}$ is fully separable.
    \item $Q=1$ for the $N$-qubit GHZ state and for any product of Bell pairs with non-trivial single-qubit rotations on each qubit.
    \item $Q$ is invariant under permutations of qubits (it is permutation-symmetric).
    \item $Q$ does not achieve its maximum on the W state, making it unsuitable as a sole target for W-state synthesis.
    \item $\mathrm{Tr}[\rho_j^2]$ can be estimated experimentally via randomised measurements \cite{MW_operationalInterpretation}, giving the measure direct experimental relevance.
\end{itemize}

\end{document}